\documentclass[prx,reprint,superscriptaddress,preprintnumbers, amsmath,amssymb,aps]{revtex4-2}

\usepackage{mathtools}
\usepackage{amsmath,bm}
\usepackage{amssymb}
\usepackage{subfigure}
\usepackage{setspace}
\usepackage{dcolumn}
\usepackage[hidelinks]{hyperref}
\usepackage{amsfonts}
\usepackage{graphicx}
\usepackage{mathrsfs}
\usepackage{times}
\usepackage{color}
\usepackage{psfrag}
\usepackage{natbib}
\usepackage{mathptmx}
\usepackage{times}
\usepackage{epstopdf}

\begin{document}

\title{Interface states in two-dimensional quasicrystals with broken inversion symmetry}

\author{Danilo Beli}
\affiliation{Department of Aeronautical Engineering, Sao Carlos School of Engineering, University of Sao Paulo, Brazil}
\author{Matheus I. N. Rosa}
\affiliation{Department of Mechanical Engineering, College of Engineering and Applied Science, University of Colorado Boulder, USA}
\author{Luca Lomazzi}
\affiliation{Department of Mechanical Engineering, Politecnico di Milano, Milano, Italy}
\author{Carlos De Marqui Jr.} 
\affiliation{Department of Aeronautical Engineering, Sao Carlos School of Engineering, University of Sao Paulo, Brazil}
\author{Massimo Ruzzene}
\affiliation{Department of Mechanical Engineering, College of Engineering and Applied Science, University of Colorado Boulder, USA}

\date{\today}

\begin{abstract}
We investigate the existence of interface states induced by broken inversion symmetries in two-dimensional quasicrystal lattices. We introduce a 10-fold rotationally symmetric quasicrystal lattice whose inversion symmetry is broken through a mass dimerization that produces two 5-fold symmetric sub-lattices. By considering resonator scatterers attached to an elastic plate, we illustrate the emergence of bands of interface states that accompany a band inversion of the quasicrystal spectrum as a function of the dimerization parameter. These bands are filled by modes which are localized along domain-wall interfaces separating regions of opposite inversion symmetry. These features draw parallels to the dynamic behavior of  topological interface states in the context of the valley Hall effect, which has been so far limited to periodic lattices. We numerically and experimentally demonstrate wave-guiding in a quasicrystal lattice featuring a zig-zag interface with sharp turns of 36 degrees, which goes beyond the limitation of 60 degrees associated with 6-fold symmetric (i.e., honeycomb) periodic lattices. Our results provide new opportunities for symmetry-based quasicrystalline topological waveguides that do not require time-reversal symmetry breaking, and that allow for higher freedom in the design of their waveguiding trajectories by leveraging higher-order rotational symmetries. 
\end{abstract}


\maketitle

\section{Introduction}\label{Introduction}
The discovery of topological insulators~\cite{hasan2010colloquium} has opened opportunities for robust wave localization and transport across multiple disciplines, including through classical waves by leveraging photonic~\cite{khanikaev2013photonic,lu2014topological}, acoustic~\cite{PhysRevLett.114.114301,ma2019topological} and elastic metamaterials~\cite{huber2016topological,miniaci2021design}. The examples realized so far illustrate a wealth of strategies for the design of backscattering-free waveguides with a high degree of immunity to defects, a feature deemed promising for technological applications and devices. Typically, these phenomena are unlocked by engineering the band structure of periodic lattices, for example by nucleating degeneracies and opening band gaps through the careful manipulation and selective breaking of their symmetries. A current challenge has been to extend these concepts to aperiodic and disordered materials in general. For example, edge states originally attributed to the quantum Hall effect (QHE)~\cite{klitzing1980new,thouless1982quantized} have been observed to persist in amorphous gyroscopic lattices~\cite{mitchell2018amorphous}. Other studies have considered quasiperiodic lattices of effective extended dimensionalities by leveraging additional dimensions established in their parameter spaces~\cite{prodan2015virtual,kraus2016quasiperiodicity}. Under this framework, edge states reminiscent of the 2D QHE have been realized in 1D quasiperiodic lattices~\cite{kraus2012topological,kraus2012topologicalb,apigo2018topological,apigo2019observation,
ni2019observation,pal2019topological,xia2020topological,marti2021edge,rosa2021exploring}, while those reminiscent of the 4D QHE have been realized in 2D quasiperiodic lattices~\cite{kraus2013four,rosa2021topologicalb,koshino2022topological}.

In this context, there has been significant interest in exploring the existence of topological states in quasicrystals (QCs). While they lack translational periodicity, QCs have long-range order~\cite{Shechtman1984,LevineSteinhardt1984} and form a particularly interesting class of aperiodic lattices since they may exhibit spatial symmetries which are forbidden in periodic crystals, such as 5, 7, 8 and 10-fold rotational symmetries in the plane, and e.g. icosahedral symmetries in 3D~\cite{walter2009crystallography}. The interplay between their unique symmetries and the concepts of topological physics is a subject of ongoing investigations~\cite{else2021quantum,fan2022topological}, which may lead to unprecedented phenomena. Some studies have demonstrated the existence of edge states reminiscent of the QHE in 2D QCs, for example in quantum lattices subject to an external magnetic field~\cite{tran2015topological,fuchs2016hofstadter,duncan2020topological}, and photonic lattices with artificial gauge fields~\cite{bandres2016topological}. The existence of higher order topological corner states have also been demonstrated in quantum QC lattices~\cite{varjas2019topological,chen2020higher}. In passive materials, which preserve time-reversal symmetry, implementations are more scarce since they must solely rely on the manipulation of their real-space symmetries. One example is the observation of topological boundary floppy modes in elastic Maxwell QC lattices~\cite{zhou2019topological}, which extends the behavior previously observed in periodic lattices~\cite{kane2014topological} to the orientation symmetries of QCs, but are limited to zero-frequency. Another example is the extension of the spin Hall effect (SHE)~\cite{kane2005quantum} to QC quantum lattices with spin-orbit coupling~\cite{huang2018quantum,huang2018theory}. While theoretically proposed, an experimental realization is still missing, perhaps due to the challenges associated with synthetizing atomic-scale quasicrystals. Additionally, the observation of the SHE in the context of classical waves is already intricate in the periodic case, where pseudo-spin degrees of freedom are generated through the careful nucleation and lifting of a double Dirac degeneracy~\cite{khanikaev2013photonic,susstrunk2015observation,mousavi2015topologically,
he2016acoustic,chaunsali2018subwavelength,miniaci2018experimental}. The absence of translational periodicity and Bloch-Floquet dispersion analysis makes the introduction of such pseudo-spin features in QCs very challenging, and, therefore, the extension of the SHE to classical QCs is still missing. 

We here investigate the existence of interface states in passive 2D QC lattices induced solely by inversion symmetry breaking. Our approach is inspired by the valley Hall effect (VHE)~\cite{xiao2007valley,drouot2020edge}, which is typically achieved by breaking the inversion symmetry of periodic 6-fold rotationally symmetric lattices (i.e., honeycomb lattice). Such framework is more straightforward than intricate designs to emulate pseudo-spin effects and provides one of the simplest avenues towards topological states in passive 2D materials, as evidenced by numerous classical wave implementations~\cite{ma2016all,noh2018observation,
lu2017observation,Torrent2013,Pal2017,vila2017observation,qian2018topology,guo2022minimal} . However, approaches based on inversion symmetry breaking have been scarcely applied to QC lattices, with current examples limited to 1D waveguides supporting interface states~\cite{apigo2019observation,davies2022symmetry}. We illustrate the phenomena using a 10-fold rotationally symmetric QC lattice whose broken inversion symmetry is achieved through a dimerization that produces two 5-fold symmetric sub-lattices (Fig.~\ref{fig:Fig1}). By considering resonators attached to an elastic plate, our simulations uncover features analogous to those observed in the context of the VHE for periodic lattices. These include a band inversion as a function of the dimerization parameter, and the appearance of interface states localized along domain walls that separate two regions of opposite inversion symmetry. We also exploit the interface states to create waveguides with sharp angular turns of 36 degrees, which goes beyond the 60 degree limitation associated with the lattice vectors of 6-fold periodic lattices. Finally, the simplicity of the proposed framework based solely on inversion symmetry allow us to experimentally demonstrate these features on a macroscopic additive manufactured elastic plate. 

This paper is organized as follows: Section~\ref{Designsec} introduces the design methodology for the dimerized quasicrystal lattices, and the implementation using elastic resonators. Section~\ref{NumericalSec} then describes the spectral properties uncovered by numerical simulations, while Section~\ref{ExperimentalSec} provides the results of experimental investigations. Finally, Section~\ref{ConclusionSec} summarizes the key findings of this study and outlines future research directions. 
  
\section{Design of Dimerized Quasicrystal Lattices With Broken Inversion Symmetry}\label{Designsec}

The considered quasicrystal lattices are obtained through the discretization of continuous fields that exhibit the desired rotational symmetry (Fig.~\ref{fig:Fig1}). A continuous field in physical space $\phi({\bf r})$, with $\bf{r}=[x,y] \in \mathcal{R}^2$, is defined by directly assigning $N$  rotationally symmetric peaks in reciprocal space ($\mathbf{k}=[k_x,k_y] \in \mathcal{R}^2$) as points in the two-dimensional Fourier spectra~\cite{Lubensky1988,Widom2008}. These peaks are angularly spaced by $\theta_N=2\pi/N$ over a circle of fundamental wavenumber $k_0 = 2\pi/\lambda_0$, where $\lambda_0$ is the fundamental wavelength. Therefore, physical and reciprocal spaces can be expressed as:

\begin{figure*}
\centering
\includegraphics[]{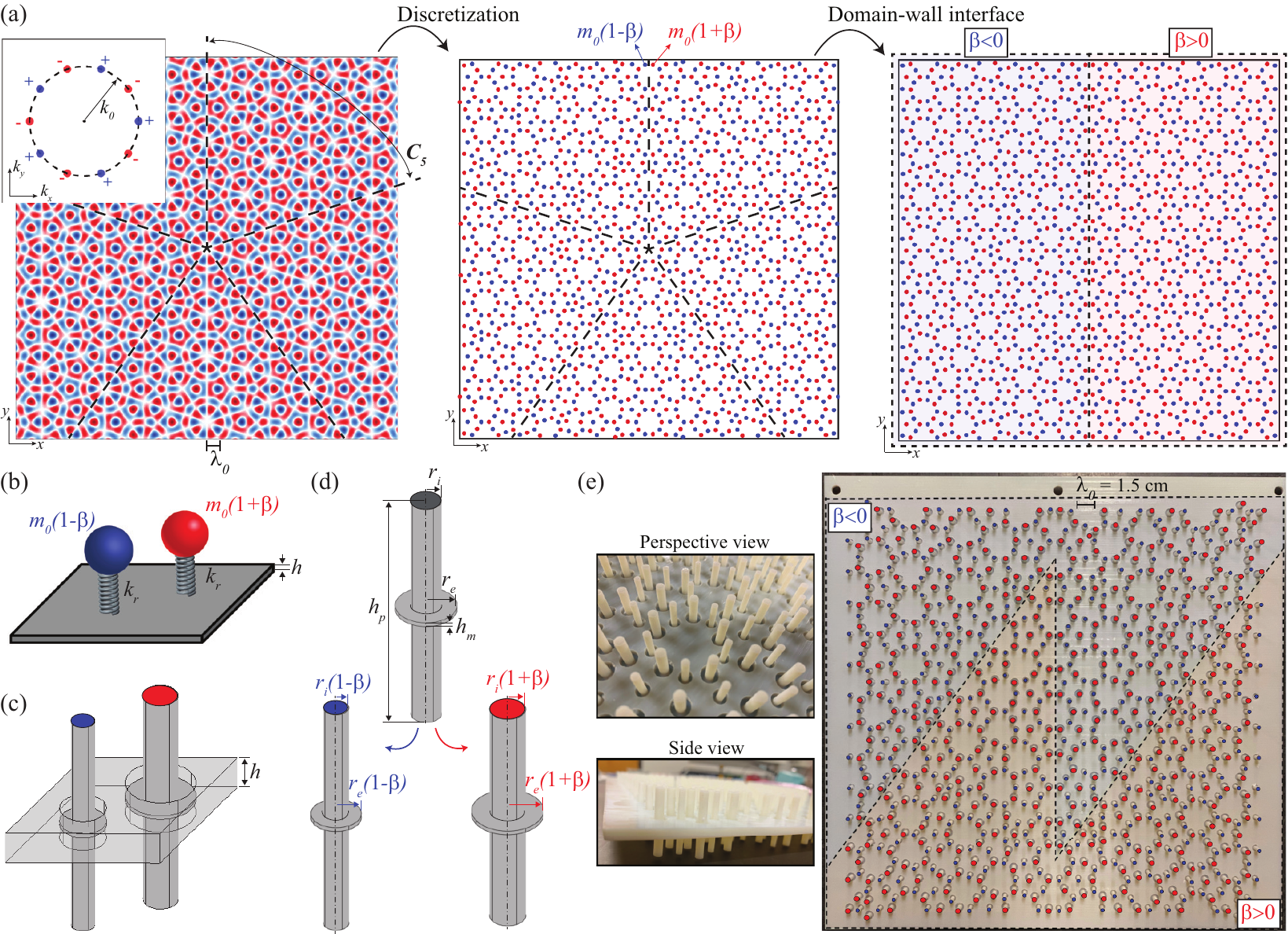}
\caption{Design of dimerized quasicrystal lattices and creation of domain-wall interfaces. (a) The quasicrystalline continuous field $\phi({\bf r})$ with $C_5$ symmetry (left panel). Inset illustrates the Fourier design peaks of alternating signs. The discretization of the field leads to the dimerized $C_5$ lattice in the middle panel, with red and blue masses parametrized by $m_0(1+\beta)$ and $m_0(1-\beta)$. A straight domain-wall interface is formed by contrasting domains with opposite $\beta$ signs, as shown in the right panel. (b) Implementation of dimerized spring-mass resonators, and (c) analogy to pillar-membrane resonators, whose dimerization illustrated in (d) is achieved via the pillars' radii. (e) Photographs of additive manufactured sample featuring a quasicrystal lattice of resonators with a zig-zag interface. The locations of bigger and smaller resonators are marked in red and blue circles overlaid to the photograph on the right.}
\label{fig:Fig1}
\end{figure*}

\begin{equation}\label{designeq}
\phi({\bf r}) = \sum_{n=0}^{N-1} e^{i ({\bf k}_n \cdot {\bf r}+\phi_n)} \text{,} \quad \hat{\phi}({\bf k}) = \sum_{n=0}^{N-1}e^{i\phi_n}\delta({\bf k}-{\bf k}_n),
\end{equation}
where $\delta$ is the delta function that locates the wave number ${\bf k}_n = k_0 [\cos \left( n \theta_N \right), \,  \sin \left( n \theta_N \right)]$ of the $n_{th}$ peak, with phase $\phi_n$. This procedure traditionally employs even $N$ and zero phases $\phi_n=0$ to produce real fields with $C_N$ symmetry (i.e. $N-$fold rotational symmetry), including crystalline (periodic) distributions, such as square and hexagonal fields with $C_4$ and $C_6$ symmetries, and quasicrystalline (quasiperiodic) distributions, such as $C_8$ and $C_{10}$ quasicrystals~\cite{Beli2021,Beli2022}. In such cases, there are $N/2$ pairs of diametrically opposite peaks of equal phases in the reciprocal space circle, and each pair defines one cosine function. The superposition of these $N/2$ cosines, which are even functions, defines a field with $C_N$ symmetry. We here introduce the phase term $\phi_n=n\pi + \pi/2$ to break the inversion symmetry of $\phi({\bf r})$, replacing $C_{N}$ symmetry by $C_{N/2}$ symmetry due to alternating phases of $\pi/2$ and $-\pi/2$ for each peak along the circle. This can be equivalently interpreted as replacing each of the $N/2$ cosine functions described above by sinusoidal ones, which are odd functions. An example of a quasicrystalline field with $C_5$ symmetry obtained with $N=10$ is shown in Fig.~\ref{fig:Fig1}(a), with the inset illustrating the Fourier peaks of alternating phases. Notice that the overall field structure resembles the $C_{10}$ symmetric case~\cite{Beli2021,Beli2022}, but the alternating blue and red colors indicate that inversion symmetry was indeed broken, characterizing the $C_5$ symmetry. Similarly, the usage of $N=6$ produces a field which breaks $C_6$ symmetry into $C_3$ symmetry, which is one of the basis for the implementation of the VHE and has inspired our design strategy. Further details on such design procedure are given in the Supplemental Materials (SM), including the periodic $C_6 \to C_3$ case, and another quasicrystalline $C_{14} \to C_7$ example.

For ease of implementation and experimentation, the continuous field is discretized to form a point lattice as illustrated in Fig.~\ref{fig:Fig1}(a). This is done through a level-cut procedure where the peaks and valleys of the continuous field respectively define two sublattices of blue and red points (see more details in the SM). The color of the points may represent a property such as mass, leading to a convenient dimerization where red and blue masses are respectively parametrized as $m_0(1+\beta)$ and $m_0(1-\beta)$, with $m_0$ being the baseline mass, and $\beta$ the dimerization parameter. Notice that the positions of the masses alone define a $C_{10}$ quasicrystalline lattice, whose symmetry is maintained when $\beta=0$ and the masses are equal to $m_0$. However, when $|\beta|> 0$, a $C_5$ quasicrystal with broken inversion symmetry is formed due to the contrast between blue and red masses. Naturally, $\beta$ values of opposite signs produce distinct versions of the $C_5$ quasicrystal with opposite inversion symmetry, which are related by interchanging the locations of red and blue masses, or by a rotation of $2\pi/10$. 

This parametrization is inspired by similar procedures applied to hexagonal lattices in the context of the VHE, where $\beta$ is used to break the inversion symmetry of $C_6$ lattices into $C_3$~\cite{pal2017edge,drouot2020edge}. Indeed, the case $N=6$ in Eq.~\eqref{designeq} reproduces the same dimerized honeycomb lattices of such prior works, as illustrated in the SM. The dimerization conveniently allows the creation of domain-wall interfaces by connecting two regions with $\beta$ values of opposite signs, as illustrated in Fig.~\ref{fig:Fig1}(a) for the case of a simple straight interface parallel to the $y$ axis. In the example of Fig.\ref{fig:Fig1}(a), the interface contrasts the red type masses by using negative and positive $\beta$ parameters respectively at the left and the right domains, while the opposite would produce an interface contrasting blue masses. More complex interfaces composed of multiple lines can be created, as long as each line is defined along one of the axis of rotational symmetry. The example in Fig.~\ref{fig:Fig1}(e) illustrates a zig-zag interface similar to those typically employed in the context of the VHE~\cite{pal2017edge}. The higher order rotational symmetry of the QCs allow for sharper turns, with the zig-zag interface presented here exhibiting two angular turns of $36^\circ$, while the periodic $C_6$ symmetry is restricted to turns of $60^\circ$. As we will show, these interfaces support localized modes with waveguiding capabilities that appear to be analogous to those found in $C_6$ symmetric periodic lattices in the context of the VHE.   

In our analysis, the proposed lattices define resonators attached to an elastic plate through springs of constant stiffness $k_r$, as conceptually illustrated in Fig.~\ref{fig:Fig1}(b). The dimerization in general leads to two distinct resonators of frequencies $\omega_a=\sqrt{k_r/m_a}$ and $\omega_b=\sqrt{k_r/m_b}$, with masses $m_a=m_0(1-\beta)$ (blue) and $m_b=m_0(1+\beta)$ (red), which are represented in Fig.~\ref{fig:Fig1}(b) in a typical honeycomb lattice unit cell. This setup is inspired by previous works dealing with elastic analogues to the VHE in elastic plates with honeycomb lattices of resonators~\cite{Torrent2013,Pal2017}. When $\beta=0$, the masses are the same and a Dirac cone is nucleated in the dispersion bands~\cite{Torrent2013}, while $|\beta|>0$ breaks the inversion symmetry of the unit cell and opens up topological band-gaps which support interface states~\cite{Pal2017}. We here extend this setup to the configurations described above based on the dimerized quasicrystalline lattices, aiming at identifying interface states with similar features. Our practical implementation consists of membrane-pillar resonators as depicted in Fig.~\ref{fig:Fig1}(c), where each resonator is composed of a thin circular membrane of thickness $h_m$ and a cylindrical pillar of height $h_p$, embedded in a baseline plate of thickness $h$. The baseline resonator is characterized by external (membrane) and internal (pillar) radii $r_e$ and $r_i$, with the dimerization leading to two distinct resonators as depicted in Fig.~\ref{fig:Fig1}(d). For example, when $\beta>0$, the smaller resonator (blue) is characterized by external and internal radii $r_e(1-\beta)$ and $r_i(1-\beta)$, while the bigger resonator (red) by $r_e(1+\beta)$ and $r_i(1+\beta)$. For the experimental part of this work, we test the QC plate shown in Fig.~\ref{fig:Fig1}(e), which features a zig-zag domain-wall interface and was fabricated through additive manufacturing (see Sec.~\ref{ExperimentalSec} for more details on the fabrication and geometrical parameters). 

\section{Spectral properties: band inversion and interface states}\label{NumericalSec}

We begin by investigating the spectral properties of the quasicrystalline plates, revealing a band inversion of the bulk bands that is accompanied by the appearance of in-gap bands of interface states. The phenomena is first numerically illustrated by considering the plate with spring-mass resonators due to the relative low computational cost and reduced wave/vibration analysis complexity. The same strategy is then implemented on the manufactured plate leveraging membrane-pillar resonators, where waveguiding along the zig-zag interface is experimentally demonstrated. The flexural motion of the thin homogeneous elastic plate is investigated based on the Kirchhoff theory, with the governing equation in the frequency domain expressed as~\cite{Torrent2013,Pal2017}:
\begin{subequations}\label{eqgov}
\begin{align}
D \nabla^4 \text{w}-\rho h \omega^2 {\text{w}} = k_r \sum_{i} \left(\text{w}-\text{w}_i \right)\delta \left(x-{\bm R}_i \right), \\
-m_i \omega^2{\text{w}}_i+k_r \left(\text{w}_i-\text{w}({\bm R}_i) \right) = 0,
\end{align}   
\end{subequations}
where $\text{w}=\text{w}(x,y)$ is the out-of-plane displacement of the plate, $\omega$ is the angular frequency, and the sub-index $i$ represents the $i^{\text{th}}$ resonator of position ${\bm R}_i$ and mass $m_i$, characterized by a displacement $\text{w}_i$. The plate is characterized by its thickness $h$, elastic modulus $E$, mass density $\rho$, Poisson ratio $\nu$, and flexural stiffness $D = Eh^3/[12(1-\nu)]$. 

\begin{figure*}
\centering
\includegraphics[]{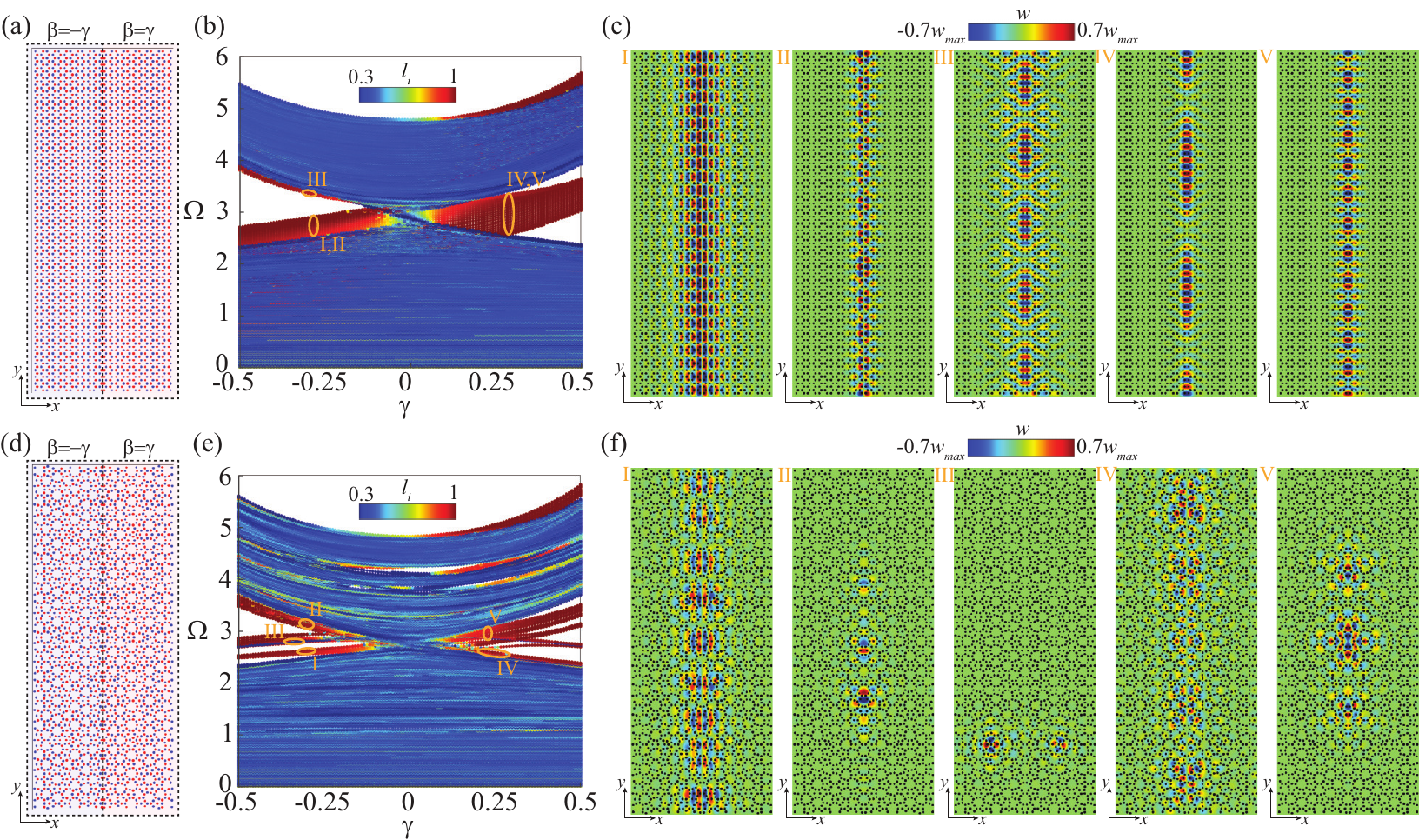}
\caption{Identification of interface states for honeycomb (a-c) and QC plate (d-f). The finite plate featuring a domain wall interface separating regions with opposite dimerization parameter $\beta=\pm \gamma$ (a,d). The spectrum as a function of $\gamma$ (b,e), color-coded by the interface localization factor $l_i$, showing the band inversion and the appearence of interface states inside the gap. Selected examples from groups of modes marked in (b,e) are displayed in (c,f).   }
\label{fig:Fig2}
\end{figure*}

Our approach relies on eigenfrequency and eigenmode computations for representative finite domains containing a straight interface, which are illustrated in Figs.~\ref{fig:Fig2}(a) and 2(d) for the periodic (hexagonal) and quasicrystalline cases, respectively. The dimerized honeycomb lattice is first presented to connect our approach and results to known features of the VHE~\cite{Pal2017}, which afterwards are analogously observed in the QC lattice. The domains are constructed by considering a plate of finite size $L_x=25\lambda_0$ and $L_y=60\lambda_0$ along the $x$ and $y$ axis, with simply supported boundary conditions at the edges ($\textup{w}=0$). The eigenfrequencies and eigenmode shapes are computed through a discretized form of Eqn.~\eqref{eqgov} solved within the Abaqus computational environment (see the SM for more details). The straight interface is introduced at the center of the domain extending parallel to the $y$ axis, separating two regions of contrasting $\beta$ values, i.e. $\beta=-\gamma$ and $\beta=\gamma$ at the right and left sides, respectively.  Two types of interfaces are possible: when $\gamma>0$, the interface contrasts resonators with the larger masses $m_0(1+\beta)$ (red), as illustrated in Figs.~\ref{fig:Fig2}(a,d), while $\gamma<0$ corresponds to an interface contrasting the smaller masses $m_0(1-\beta)$ (blue). Varying $\gamma$ therefore allows for a transition between these two configurations, with the transition point $\gamma=0$ defining a lattice with no interface since all masses are equal to $m_0$.

The results in Figs.~\ref{fig:Fig2}(b,e) show the variation of the eigenfrequencies with $\gamma$ for the hexagonal and quasicrystal lattices, respectively. For ease of comparison with previous studies~\cite{Torrent2013,Pal2017}, our results are presented in terms of the normalized frequency $\Omega=\omega a^2 \sqrt{\rho h/D}$, where {$a= 2\sqrt{3}\lambda_0/3$} is the resulting hexagonal lattice constant when $N=6$ is used in Eqn.~\eqref{designeq}. We also employ a baseline resonator frequency $\Omega_0=4\pi$, which sets $k_r=16\pi^2 m_0 D/\rho h a^4$, and a mass ratio of $10$, i.e. $m_0=10\rho A_c h$, where $A_c= \sqrt{3}a^2/2$ is the area of the honeycomb lattice unit cell. The eigenfrequencies are color-coded according to an interface localization factor $l_i$, which is defined as: 
\begin{equation}\label{eqlocfactor}
l_i=\dfrac{\int_{\mathcal{D}_i}  \text{w}^2 dA }{\int_{\mathcal{D}} \text{w}^2 dA },
\end{equation}
where $\text{w}$ is the mode shape, $\mathcal{D}$ represents the domain of the plate, and $\mathcal{D}_i$ denotes a smaller domain centered near the region of the interface with $40\%$ of the length $L_x$, extending uniformly along the $y$ direction. High $l_i$ values indicate eigenfrequencies whose eigenmode shapes are concentrated near the interface (red), while low $l_i$ values correspond mostly to non-localized bulk modes (blue). The spectra also typically exhibit modes localized at the edges of the plate, which were filtered out from the plots based on a similar localization factor defined for the edges (see SM). 

Both the honeycomb and QC plate spectra in Fig.~\ref{fig:Fig2} feature a large band-gap, defined by the absence of bulk (blue) modes, that undergoes a band inversion as it closes and re-opens as a function of $\gamma$. This behavior is commonly found in the search for topological states, usually defining phases of different topological properties, and therefore is a strong indicator that topological states will be found when contrasting domains living in opposite sides of the band inversion~\cite{lu2017observation,guo2022minimal}. Indeed, due to the presence of the domain wall, we observe that additional bands of modes localized at the interface (red) appear inside the band-gaps. These features are reminiscent of those observed in the context of the VHE, and, in the honeycomb lattice, they can be precisely mapped to previously established results. For instance, the gap closing point at around $\Omega=2.74$ for $\gamma=0$ in Fig.~\ref{fig:Fig2}(b) corresponds to the Dirac cone degeneracy identified by the Bloch analysis of the hexagonal unit cell~\cite{Torrent2013}. The degeneracy is lifted for $|\gamma|>0$ due to the inversion symmetry breaking, opening up a band-gap with non-trivial topological properties related to Valley Chern numbers of opposite signs at the valley $K$ and $K'$ points~\cite{Pal2017}. By contrasting two domains with opposite $\beta$ values, and therefore opposite valley Chern numbers, the existence of interface states is typically identified by considering a $x-$wise finite strip and imposing periodicity conditions along the $y$ direction. In Fig.~\ref{fig:Fig2}(b), the bands of interface states contain a finite number of modes that are a result of the finite extent along the $y$ direction, which causes a discretization of the otherwise continuous bands in the infinite case. Selected mode shape examples from the groups of bands highlighted in Fig.~\ref{fig:Fig2}(b) are displayed in Fig.~\ref{fig:Fig2}(c). In the left half of the spectrum ($\gamma<0$), there are two groups of modes inside the gap, one at its lower and the other at its upper boundary, containing modes which are anti-symmetric (I,II) and symmetric (III) with respect to the interface line, respectively. In the right half ($\gamma>0$), there is one group of modes inside the gap which are symmetric with respect to the interface (examples IV,V). The correspondence between these results and the analysis typically conducted for periodic lattices based on Bloch-Floquet theory is illustrated through additional numerical results in the SM. However, such periodicity conditions cannot be applied in the QC case, which motivates our approach based on the large finite domains. The QC spectrum in Fig.~\ref{fig:Fig2}(e) exhibits a similar band inversion with a gap closing point at $\gamma=0$ around the Dirac cone frequency of the honeycomb lattice. The interface states form a larger number of thinner bands when compared to the hexagonal lattice, with selected modes displayed in Fig.~\ref{fig:Fig2}(f) exemplifying their characteristics. A band of anti-symmetric modes (I) is present at the lower boundary of the gap for $\gamma<0$, and migrates to the upper boundary of the gap (V) for $\gamma>0$. Similarly, a band of symmetric modes migrate from the upper boundary (II) to the lower boundary (IV) of the gap as $\gamma$ increases and flips sign. While these modes are localized at the interface, in contrast to the hexagonal case they are not necessarily extended throughout the entire interface due to the absence of periodicity (see for example modes II and V). The QC also exhibits a thin band of modes inside the gap which are localized in isolated regions away from the interface, as exemplified by mode III. These are reminiscent of intrinsic localized modes found in other investigations of QC lattices~\cite{wang2003localized,della2006localized,mnaymneh2007mode,sinelnik2020experimental}, which are expected to appear even in the absence of defects due to the lack of translational periodicity. More details on the in-gap states of the QC plates are provided through additional mode plots in the SM.

\begin{figure}
\centering
\includegraphics[]{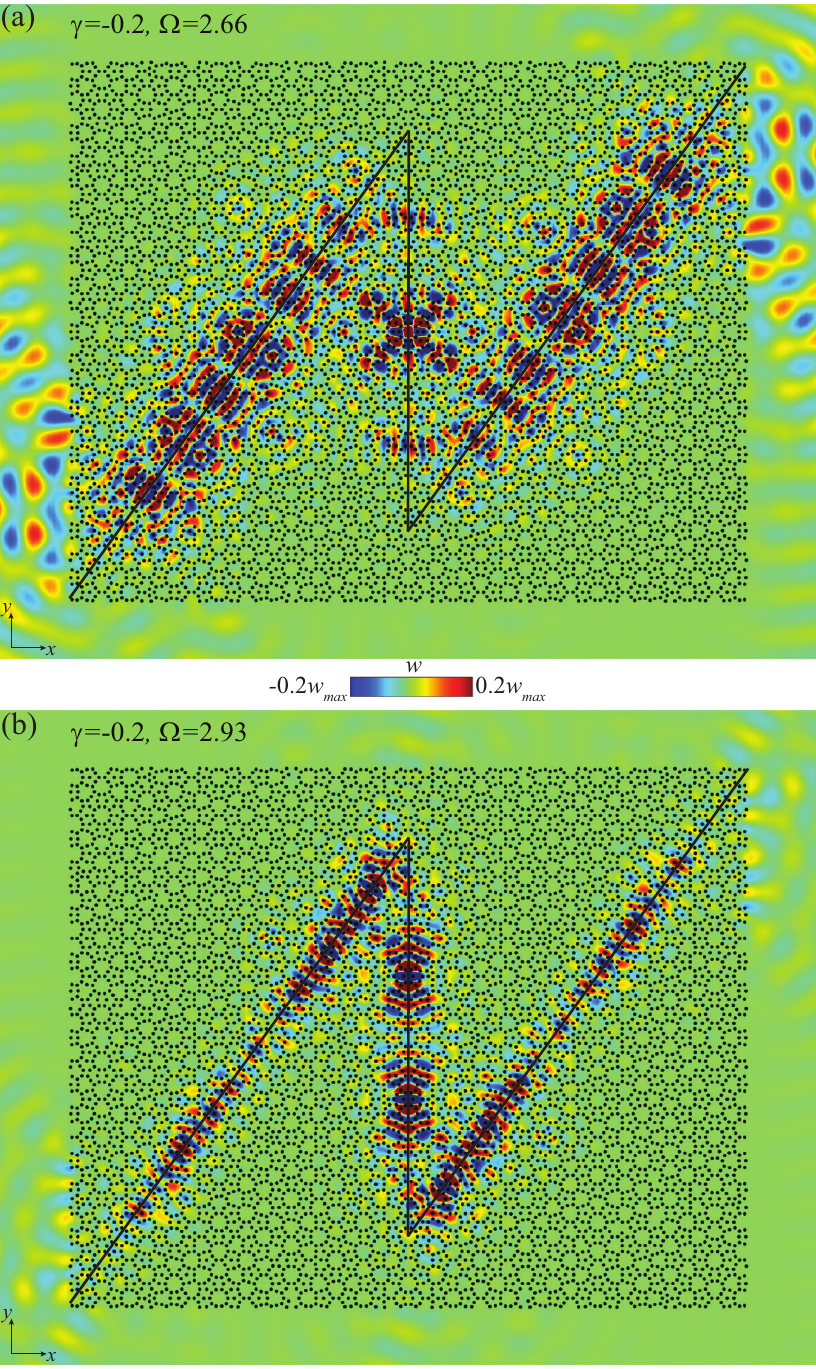}
\caption{Harmonic response of QC plate with zig-zag interface for selected frequencies, illustrating the excitation of anti-symmetric (a) and symmetric (b) interface states that propagate through angular turns of $36^\circ$.}
\label{fig:Fig3}
\end{figure}

Our results illustrate the emergence of interface states due to the broken inversion symmetry of the QC lattice, in analogy to the behavior of hexagonal lattices in the VHE. The finite elastic plate analysis provides a prediction of the frequency ranges occupied by the interface states, and their features such as symmetric vs anti-symmetric mode shapes. These modes can be exploited to create waveguides with sharp turns that leverage the symmetry axes of the resonator lattice. To exemplify,  Fig.~\ref{fig:Fig3} displays the response of a QC plate with $\gamma=-0.2$ and a zig-zag interface for two frequencies that excites the bands (I) and (II) of Fig.~\ref{fig:Fig2}(e). The harmonic response to a point source placed at the center of the plate is computed for the selected frequencies through a multiple scattering procedure, which is detailed in other works~\cite{Torrent2013,marti2021edge}. We observe that the displacement field is localized at and extends throughout the entire interface, with no evident backscattering at the angular turns. The results are also in agreement with the predicted features of the interface states, with Fig.~\ref{fig:Fig3}(a) and (b) respectively showing anti-symmetric and symmetric modes with respect to the interface line, corresponding to the expected behavior of bands (I) and (II) from Fig.~\ref{fig:Fig2}(e).

\section{Experimental Observation of Interface States in Quasicrystal Elastic Plate}\label{ExperimentalSec}

The physical demonstrator in Fig.~\ref{fig:Fig1}(e) was fabricated through additive manufacturing due to its geometrical complexity. Specifically, the fabrication was done in a Stratasys PolyJet machine using the Vero Rigid material, with estimated properties: elastic modulus $E = 3$ GPa, mass density $\rho = 1000$ kg/m$^3$, and Poisson coefficient $\nu = 0.33$. The zig-zag domain-wall was chosen to confirm the wave propagation of the interface state in the presence of very sharp angular turns of 36$^o$. The position of the resonators were obtained from the described design procedures with a fundamental length scale of $\lambda_0=1.5$cm, and the plate has a total size of $39 \times 39$cm. The plate is characterized by the following geometrical parameters: $h = 4$ mm, $h_{p} = 32$ mm, $h_{m} = 0.5$ mm, $r_{e} =3.3$mm, $r_{i} = 1.65$mm, with a dimerization parameter of $\beta = 0.15$. These parameters were chosen guided by 3D finite element simulations to establish the features of the VHE in the periodic (hexagonal) lattice, which are detailed in the SM. In particular, when $\beta=0$ we achieve a Dirac cone degeneracy at the frequency $f=6915$ Hz, while $\beta=0.15$ lifts the degeneracy and opens a band-gap that extends from $f=5822$ Hz to $f=7548$ Hz. Our simulations then confirm the existence of interface states in the QC plate with the zig-zag interface at the same frequency range, including a frequency response and a time transient analysis that are also shown in the SM, while here we focus on experimental results.

\begin{figure*}
\centering
\includegraphics[]{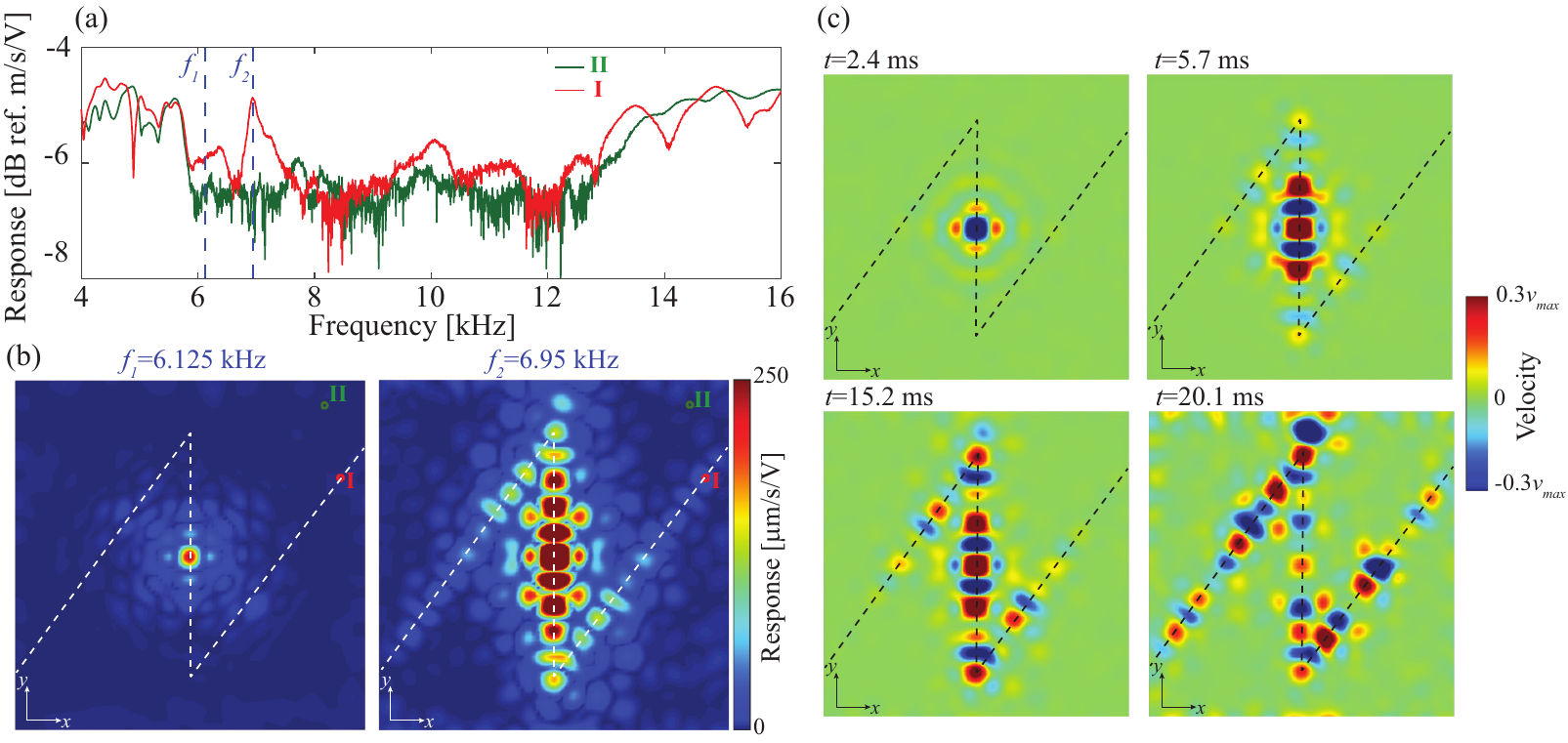}
\caption{Experimental observation of interface states in QC additive manufactured plate. The frequency response of the plate for points I and II (a), which are marked in the response fields for selected frequencies $f_1$ and $f_2$ displayed in (b). The resonance peak inside the gap identified at $f_2$ is characterized by a response concentrated along the zig-zag interface. The snapshots for subsequent time instants in (c) illustrate the transient wave propagation along the interface caused by a wavepacket centered at $f_2$.}
\label{fig:Fig4}
\end{figure*}

In experimental tests, the out-of-plane motion of the plate (i.e., $z-$axis) is measured by a Scanning Laser Doppler Vibrometer (SLDV). An excitation is provided through a piezoelectric disc placed at the center of the plate (STEMINC model SMD05T04R411). The excitation signal is generated by the computer and amplified by a Linear Amplifier (PIEZO SYSTEMS EPA-104) before reaching the piezo disc. The motion is recorded across a square grid containing $90$x$90$ points, and the results are combined and processed within the Matlab environment. First, a broad-band sinusoidal chirp signal in the frequency range of $4-16$ kHz is used to characterize the frequency response of the plate, which is estimated by taking the ratio between the Fourier Transforms of the measured velocity and the input voltage. Figure~\ref{fig:Fig4}(a) displays the response at two points, one at the interface (I), and the other at the corner of the plate (II). A large band-gap extending approximately from $6$ kHz to $12$ kHz is observed to appear, due to the broken inversion symmetry from the dimerization with $\beta=0.15$. While the response near the corner (II) evidences an almost uniform gap, the response at the interface (I) highlights an additional resonance peak residing within the gap at $6950$Hz. The response for the entire played is displayed for two selected frequencies in Fig.~\ref{fig:Fig4}(b): at $f_1=6125$ Hz a typical spatial attenuation occurring inside the gap is shown, while for the resonance peak at $f_2=6950$ Hz the response is clearly localized along the zig-zag interface. The excited interface state exhibits a symmetric shape similar to those of type V in Fig.~\ref{fig:Fig2}(f). We also note that there is a significant, albeit uniform, amplitude decay away from the center of the interface. We attribute this behavior to the high dissipation inherently present in the 3D printed material. This statement is also supported by additional numerical simulations presented in the SM, where a comparison between the simulated responses with and without dissipation is shown. 

After the experimental characterization of the frequency spectra and identification of the interface state frequency, we do an additional transient test to illustrate the temporal wave propagation along the zig-zag interface of the QC plate. A sinusoidal wavepacket centered at the frequency $f=6950$ Hz with 100 cycles  is sent to the piezoelectric patch, with the response simultaneously recorded by the SLDV. The panels in Fig.~\ref{fig:Fig4}(c) show subsequent snapshots of the recorded wave-field, with a video animation of the entire response included in the SM. Due to the high dissipation of the material, the color axis is separately scaled for each snapshot in a range corresponding to $30\%$ of the point with the maximum recorded velocity $v_{max}$ at that time instant. The results clearly show the waves propagating along the interface with no evident backscattering, and successfully passing through the angular corners of $36^\circ$, thus confirming the expected features from the interface states in the QC plate induced by the broken inversion symmetry.  

\section{Conclusions}\label{ConclusionSec}

This paper proposes a framework that extends the mechanism of inversion symmetry breaking and its utilization for the creation of domain-wall interfaces to QC lattices. Inspired by the features of the VHE in periodic lattices, our results confirm the existence of a band inversion accompanied by the appearance of interface states upon varying a dimerization parameter. Both numerical and experimental results illustrate possibilities for wave-guiding along interfaces with sharper turns enabled by the higher-order rotational symmetries of QCs, thus extending the behavior previously restricted to symmetry orders of periodic lattices. While these features are illustrated for elastic waves, they can be more generally applied to other types of waves including acoustic, electromagnetic, and even to quantum systems. Indeed, our results pave the way for the exploration of topological states in QCs through a simple inversion symmetry breaking procedure, which does not require magnetic fields or active components to break time-reversal symmetry. Although our results demonstrate a strong correlation to the VHE both through numerical and experimental results, future efforts should be dedicated towards a rigorous understanding of the underlying topological properties and features that are challenging to explore in the absence of Bloch-Floquet periodicity analyses. These may include the observation of momentum space features such as the Dirac cone, the localization of the Berry curvature, and computation of topological invariants such as the Valley Chern number, or any equivalent features that are also present in the quasicrystalline case. On the practical side, our results show strong evidence of backscattering-free wave propagation, but a more rigorous analysis and quantification of backscattering alongside comparisons with trivial or deffect-based waveguides is an important task for future studies.~\cite{jin2018robustness,orazbayev2019quantitative,arregui2021quantifying}.

\begin{acknowledgments}
D. Beli and M. I. N. Rosa contributed equally to this work. D. Beli and C. De Marqui Jr. gratefully acknowledge the support from S\~{a}o Paulo Research Foundation (FAPESP) through grant reference numbers: 2018/18774-6, 2019/22464-5 and 2018/15894-0 (Research project - Periodic structure design and optimization for enhanced vibroacoustic performance: ENVIBRO). M. I. N. Rosa and M. Ruzzene gratefully acknowledge the support from the National Science Foundation (NSF) through the EFRI 1741685 grant and from the Army Research Office through grant W911NF-18-1-0036.
\end{acknowledgments}

\bibliographystyle{unsrt}
\bibliography{References}
\end{document}